\documentclass{emulateapj}

\usepackage{amssymb}

\slugcomment{Accepted in The ApJ - May 6, 2004}

\shorttitle{Interaction/Merger Rate based on Collisional Ring Galaxies}
\shortauthors{Lavery et al.}

\begin{document}


\title{Probing the Evolution of the Galaxy Interaction/Merger
Rate Using Collisional Ring Galaxies}

\author{Russell. J. Lavery}
\affil{Department of Physics \& Astronomy, Northern Arizona University, Flagstaff, AZ 86011}
\email{Russell.Lavery@nau.edu}

\author{Anthony Remijan\altaffilmark{1,2}}
\affil{Department of Astronomy, University of Illinois, Urbana, IL 61801}
\email{aremijan@astro.uiuc.edu}

\author{Vassilis Charmandaris\altaffilmark{3}}
\affil{Astronomy Department, Cornell University, Ithaca, NY 14853}
\email{vassilis@astro.cornell.edu}

\author{Richard D. Hayes\altaffilmark{4}}
\affil{Department of Physics \& Astronomy, Northern Arizona University, Flagstaff, AZ 86011}


\author{Amy A. Ring}
\affil{Department of Mathematics, Northern Arizona University, Flagstaff, AZ  86011}


\altaffiltext{1}{Current address:  NASA Goddard Space Flight Center, 
Earth and Space Data Computing Division, Code 930, Greenbelt, MD 20771}

\altaffiltext{2}{National Research Council Resident Research Associate}
\altaffiltext{3}{Chercheur Associ\'e, Observatoire de Paris, LERMA, 61 Av. de l'Observatoire, F-75014 Paris, France}

\altaffiltext{4}{Deceased, April 3, 2003}


\begin{abstract}

We present the results from our program to determine the evolution
of the galaxy interaction/merger rate with redshift using the 
unique star-forming characteristics of collisional ring galaxies.
We have identified 25 distant collisional ring galaxy candidates
(CRGCs) in a total of 162 deep {\em Hubble Space Telescope} Wide
Field/Planetary Camera-2 images obtained from the {\em HST}
Archives.  Based on measured and estimated redshifts, these
25 CRGCs all lie in the redshift interval of 0.1 $\le z \le$ 1.
Using the local collisional ring galaxy volume density and the
new ``standard'' cosmology, we find that in order to account for
the number of identified CRGCs in our surveyed fields,
the galaxy interaction/merger rate, parameterized as $(1 + z)^{m}$,
must increase steeply with redshift.
We determine a minimum value of $m$ = 5.2 $\pm$ 0.7, though $m$
could be as high as 7 or 8.
We can rule out a non-evolving ($m$ = 0) and weakly evolving
($m$ = 1-2) galaxy interaction/merger rate at greater than
the 4 sigma level of confidence.

\end{abstract}



\section{Introduction}

It is becoming increasingly evident that the interactions between and
the merging of galaxies have contributed substantially to the
evolution of galaxies, both in terms of their stellar populations and
their morphological appearances.  Many of the nearby elliptical
galaxies we observe in the general field may have been formed as a
result of the merging of two large disk systems
\citep{Toomre72,Toomre77,Schweizer83,Schweizer92}.
The merging of galaxies may also be responsible for the
disappearance of the large faint blue galaxy population found
in redshifts surveys and galaxy
count programs \citep{Carlberg92,Broadhurst92,Colin94}.
And while galaxy interactions and mergers have been associated with
large bursts of star formation, there is also evidence that
the lack of interactions may also affect the star formation
processes in galaxies, as evidenced by the correlation between
the lower rates of star formation with fewer companion galaxies
found for low surface brightness galaxies \citep{Bothun93}.

Observationally, there is at least qualitative evidence for an
increase in the galaxy interaction rate with redshift based on the
increase of several phenomena accepted as resulting from interactions
between galaxies.  These include the starburst IRAS galaxies
\citep{Lonsdale90} and quasars \citep{Boyle88}.  Additionally, in the
environs of distant rich clusters of galaxies, the presence of an
excess of blue star-forming galaxies in these clusters, known as the
``Butcher-Oemler'' effect \citep{Butcher78,Butcher85}, is at least
partially, if not totally, the result of interacting galaxy systems
\citep{Lavery88,Lavery92b,Couch94,Dressler94}.

Several theoretical arguments have also been put forth supporting a
steep evolution of the galaxy merger rate as a function of redshift.
The parameterization of this evolution is expressed as $(1+z)^{m}$.
\citet{Toomre77} considered two-body encounters in an
$\Omega$ = 1 universe and suggested
that $m$ = 2.5. \citet{Carlberg90}
has suggested that the value of $m$ is sensitive to the value of
$\Omega$.  Assuming a CDM-like cosmology, \citet{Carlberg90}
has derived the following relationship:

$$ m = 4.51\,\Omega^{0.42}. $$

For $\Omega$ = 1, a very steep increase in the interactions
and mergers of galaxies with redshift is predicted and
this expectation can be tested observationally.

Unfortunately, a direct measure of $m$ has proven to be quite
difficult.  The main challenge in quantitatively determining
the increase of interactions and mergers with redshift is the
difficulty of identifying a complete volume sample of such
systems, even at moderate redshift.
Since tidal features and distortions
are, in many cases, weak and of low surface brightness,
it is often problematic to observe such features in high
redshift galaxies \citep[see][]{Mihos95,Hibbard97}.

Due to the problems associated with simply using the disturbed
morphological appearance of a galaxy to identify it as
having undergone a
recent interaction, the observational programs to investigate
the evolution of the galaxy interaction rate have been based
on determining the evolution of the
galaxy pair fraction as a function of magnitude and/or redshift.
The evolution of the galaxy pair fraction as a function of
redshift has been parameterized as $(1 + z)^{n}$, similar in
form to the galaxy merger rate with an exponent $n$ rather
than $m$.

\citet{Zepf89} first applied this method to determine this
exponent $n$ using deep multi-color plates taken on the KPNO 4-meter Mayall
telescope.  Based a complete magnitude limited sample of $\sim$1000
galaxies, they found an excess of close pairs of galaxies consistent
with a value of $n$ equal to 4.0 $\pm$ 2.5.
A similar approach using
images obtained on the Canada-France-Hawaii Telescope has been
undertaken by several investigators.  \citet[][hereafter
CPI]{Carlberg94}, determined a value of $n$ = 2.3 $\pm$ 1.0
analyzing a sample of $\sim$400 galaxies.
\citet{Yee95} using a sample of $\sim$100 field galaxies with
measured redshifts determined a value of $n = 4.0 \pm 1.5$.
But, in opposition to these determinations of \*n
is the result of \citet{Woods95}.
Their magnitude limited sample of $\sim$900 galaxies was
complete at a level of two magnitudes below that of the two
previously mentioned CFHT programs, yet
\citet{Woods95} found no evidence for any
evolution of the galaxy pair fraction ($n$ = 0).  Finally, in the
study of 545 field galaxies with an average
redshift of $<z>$ = 0.33
from the CNOC survey on the Canada-France-Hawaii Telescope,
\citet{Patton97} determined an intermediate value for $n$, with $n =
1.8 \pm 0.9$.

This significant variation in the value of $n$ cannot be simply
attributed to differences in the image resolution of the
various data sets.
Analysis of a sample of 146 galaxies in a field imaged with the
{\em HST} by \citet{Burkey94}, complete to the same magnitude
limit as that of \citet{Woods95}, resulted in a relatively
large value for $n$, with $n = 3.5 \pm 0.5$.
Yet, from the Medium Deep
Survey {\em HST} Key project,
\citet{Neuschaefer95}, using a sample of
$\sim$4500 galaxies, found no excess of galaxy pairs,
consistent with
a non-evolving ($n$ = 0) galaxy pair fraction out to a redshift of
0.5.  Interestingly, while CPI and \citet{Neuschaefer95} have
determined very different values for $n$, $2.3\pm1.0$ and $\sim$0
respectively, they have both determined the
observed faint galaxy pair
fraction to be $\sim$14\%!  The difference in their values for $n$
results from the determination of the correction for non-physical
galaxy pairs. with CPI estimating $\sim$4\% to be non-physical pairs
and \citet{Neuschaefer95} estimate the correction to be 13\%.

Besides the difficulty in determining the value of $n$ for the
evolution of the galaxy pair fraction, relating $n$ to the exponent
used for the evolution of the galaxy merger rate, $m$, is not
straightforward, as it 
is {\em extremely difficult to estimate what fraction of the galaxy 
pairs which may show a disturbed morphology will actually lead to a 
violent interaction of a  merger}.
\citet{Burkey94} suggest the $z$ derivative of the
pair fraction should represent the galaxy merger rate,
meaning $m = n - 1$.
But, if the ratio of merger timescales at different epochs is
taken into account, as was done by CPI,
the merger rate exponent, $m$,
is approximately equal to $n + 1$. \citet{Zepf89} and \cite{Yee95}
argue that the evolution of the galaxy merger rate should be quite
similar to that of the galaxy pair fraction, implying $m = n$.

     Because of these difficulties with the analysis of distant
galaxy pairs, several research groups have utilized the 
high-resolution imaging capabilities of the {\em HST}
to quantify the galaxy merger rate at high redshift.  
Based on a sample of 285 galaxies with redshifts out to $z$ = 1,
\citet{Lefevre00} used visual classifications of galaxies along
with a galaxy pair analysis to determine a value for $m$ of
3.4 $\pm$ 0.6 (for comparison with more recent results,
it should be noted that this value was determined
assuming $q_0$ = 0.5 and would be lower in the new ``standard''
cosmology).
For even higher redshifts, \citet{Conselice03}
have used a statistical measure of the asymmetry of a galaxy to
quantify the number of galaxy mergers as a function of redshift.
Using galaxies in the {\em Hubble Deep Field}, they find a steep
evolution of the galaxy merger rate for massive, luminous
galaxies out to $z$ $\sim$ 3, with a value for $m$ in the range
of 4 to 6.  The importance of the image resolution provided by
the {\em HST} cannot be understated and we have also utilized
this capability to investigate distant collisional ring galaxies.

Collisional ring galaxies (CRGs) are a relatively small fraction
of all galaxies that have undergone a recent interaction,
being produced in
the relatively rare circumstance of a small galaxy passing directly
through the center of a disk galaxy.  The ``Cartwheel'' galaxy
\citep{Zwicky41} is probably the most
well-known example of this type
of galaxy.  A thorough review of the properties of collisionally
produced galaxies is given by \citet{Appleton96}.  While the ring
structure of these galaxies identifies them as galaxies having
undergone a recent interaction, at redshifts of $z$ $\sim$ 0.5,
these galaxies will be only several arcseconds in diameter
and difficult to
identify with even the best image resolution obtainable with
ground-based observatories.

Collisional ring galaxies have several advantages for investigating
the evolution of the galaxy interaction rate:

\begin{enumerate}

\item The signature morphological appearance of CRGs, a high
surface brightness knotty ring, makes it possible to identify
them at high redshift ($z \le 1$) given the high image
resolution obtainable with the {\em HST}.

\item The ring structure, delineated by regions of massive star
formation, is a {\em direct} result of the dynamical process of
interest.  The uncertainty in how $m$ and $n$ are related is
avoided.

\item The timescale for the presence of the ring structure
is short (1 or 2 dynamical times), meaning the interaction
was a recent event.

\item The ring structure will only be produced
by an ``intruder'' galaxy
that has a significant mass compared to the ``target'' galaxy (at
least 10\%).  Therefore, only events that will have a significant
effect on the morphological appearance of the galaxy are
investigated.

\end{enumerate}

In this paper, we present the results from our program to
identify distant collisional ring galaxy candidates (CRGCs)
in deep WFPC2 images obtained from the {\em HST} Archives.
Our search method and results, the identification of 25 distant
CRGCs, and the spectroscopic observations of several of these
CRGCs are presented in Section 2.
In Section 3, we discuss the implication of these results on
the redshift evolution of the galaxy interaction/merger rate,
with Section 4 being a discussion of corrections and possible
biases affecting our value for $m$.
Our conclusions, consistent with the ``work-in-progress''
results presented in \citet{Lavery00}, 
are summarized in Section 5.

\section{Observations}\label{sect_obs}
\label{sec:obs} 

\subsection{ {\em Hubble Space Telescope} Data}

The concept for this program originated from the results of
\citet{Lavery96} who found a high density of distant CRGCs
lying behind the Local Group dwarf spheroidal galaxy in
Tucana \citep{Lavery92}.
The data of \citet{Lavery96} consisted of two deep images,
with total exposure
times of 3.5 and 4 hours, in the broad {\em HST} F814W filter.
With these long exposure times in mind, our initial selection of
WFPC2 fields to be searched for CRGCs were pointed observations
(PIs:  Couch, Dressler, Ellingson, Miley and Stockton) of distant
clusters of galaxies and distant radio galaxies which were obtained
from the {\em HST} Archives.  The only constraint, besides exposure
time, on the field selection was that at least one set of
observations were made through one of the broad {\em HST} $R$-band
or $I$-band filters (F606W, F622W, F702W, F814W).
This filter constraint was chosen so that
the identification of the CRGCs would be based on their
appearance in the rest-frame $B$-band.
The {\em Hubble Deep Field} was also included in this sample.
These pointed observations provided a total of 43 WFPC2 fields.

Based on the experience gained from the analysis of the pointed
observations, we found that fields with exposures times greater
than 4000 seconds would have a sufficient signal-to-noise ratio
for the classification of distant galaxies as CRGCs.
With this constraint,
a total of 118 WFPC2 fields were obtained from Parallel Observation
Programs (PI: Griffiths, Windhorst).  These fields, along with
the pointed observations, provided a sample of 162 WFPC2 fields
in total.

The identification of the CRGCs was done through visual inspection
of the processed frames (after cosmic ray removal and
image addition).
Each field was searched independently by some combination of
three of the authors (RJL and two others).
After completion of the independent identifications, the candidate
lists were compared and a consensus was reached on those
objects to be included in our final CRGC list.

Visual inspection was done rather than employing an automated
identification routine for several reasons.  First, the number of
objects being searched for is very small compared to the total number
of objects that would need to be structurally analyzed.  Roughly
expecting about 1 CRGC in every 10 WFPC2 fields, several
hundreds of galaxies would have to be analyzed.  While distinguishing
between exponential disks and de Vaucouleurs profiles can be done,
the construction of algorithms for the identification of coherent
ring structure is much more complex.
Second, it would still be necessary
for visual classification to distinguish the P-type and O-type rings
and to determine the presence of any weak bar-like or spiral
structure.  Additionally, the reimaging optics in WFPC2 camera
system produce ring shaped artifacts due to bright stars in
the field of observation.
These ring artifacts have a knotty structure and look like face-on
empty collisional ring galaxies.  Their location with respect to the
bright star varies in both the distance from the star and
its angular
position, depending on the location of the star on the CCD chip.
But, fortunately, these ``ringers'' can be identified as
artifacts as a line drawn
from the center of the CCD through the bright star will bisect the
ring-shaped image produced by that star.

To provide consistency between the classifiers in the visual
identification process, the ``reclassification'' of the
collisional ring galaxies in the Few \& Madore
(1986, hereafter FM86) sample of ring galaxies
was independently done by the classifiers prior to inspecting the
WFPC2 fields.  The FM86 sample of ring galaxies contains 69 objects
consisting of two types: O-type or resonant rings and P-type or
collisional rings.  Resonant or O-type rings, which comprise
$\sim$40\% of the FM86 sample, have a smooth structure and a
centrally located nucleus.  The P-type or collisional rings, which
exhibit a ring with a crisp knotty structure and sometimes have a
displaced nucleus,
constitute the remaining $\sim$60\% of the FM86 sample.
A total of 50 galaxies were selected,
consisting of the 41 P-type ring galaxies in the FM86 sample
along with 9 of their O-type ring galaxies, and
were ``reclassified'' independently and without prior knowledge of
the FM86 classifications, by each visual inspector using extracted
images from the Digitized POSS.
For the 9 O-type rings in this sample,
there was strong agreement with the FM86 classification.
But, of the 41 P-type rings, unanimous agreement with the P-type
ring classification was reached for only 14
galaxies, with 2/3 of the galaxies not being classified as
P-type by the authors of this paper.
Most of the discrepancy was due to either the presence of
spiral structure, which produced a ring-like appearance,
or the presence of a galactic bar which has induced a ring-like
structure more consistent with the O-type ring morphology.

\begin{figure*}[!ht]
\figurenum{1}
\plotone{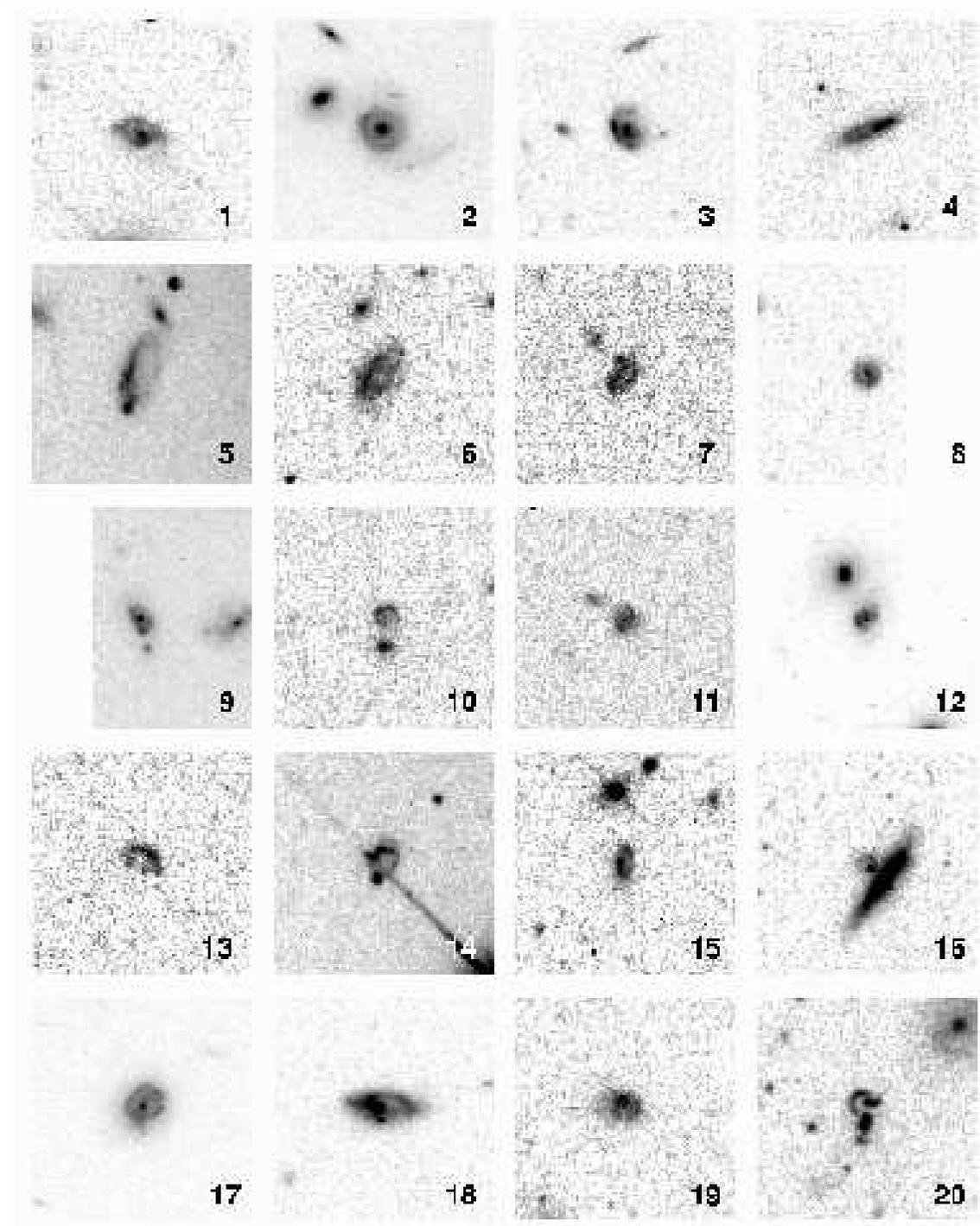}
 
\caption{ {\em HST} WFPC2 images of 20 of our identified CRGCs.
Each galaxy is identified by an index number at the bottom right
corner as given in Table 1.  These images were extracted from
the broad {\em HST} $R$-band or $I$-band
(F606W, F622W, F702W, F814W) data sets used in the
identification process.
Each image in the panel is 10 by 10 arcseconds. }

\end{figure*}

We reached two important conclusions from this reclassification
exercise.  First, our classification of galaxies as P-type rings
is very conservative.  This gives us confidence in our ability to
identify CRGCs that have a high likelihood of being true CRGs
implying that our sample will not be contaminated by
misclassified ``normal'' galaxies.
Second, based on our reclassification of the FM86 sample, in
which we found only 14 of the 41 to meet our criteria for being
P-type ring galaxies, the local
CRG volume density may only be 1/3 of that determined by FM86.

The result of our visual inspection of 162 WFPC2 fields is a total of
25 distant CRGCs, for which images of 20 are presented in Figure 1.
The photometric properties of our 25 CRGCs is presented in Table 1.
For the photometric calibration, we have adopted the standard
photometric zero-points and color coefficients determined by
\citet{Holtzman95}.  As a number of our CRGCs were identified in
Parallel fields, they were only observed in a single filter.  Eight of
our CRGCs were observed in two filters allowing for color measurements
for which we find an average ($V$-$I$) color of 1.59 $\pm$ 0.15.  As
we will be using $V$-band magnitudes of the CRGCs to estimate their
redshifts, for those CRGCs with only a single observation, we have
adopted colors of ($V$-$I$) = 1.5 and ($V$-$R$) = 1.0 in order to
determine standard $V$-band magnitudes.  These magnitudes should be
accurate to $\pm$0.1 mag.

\begin{deluxetable*}{cccccccccccc}
\tabletypesize{\scriptsize}
\tablecaption{Observed Properties of Collisional Ring Galaxies Candidates (CRGCs)\label{tbl1}}
\tablewidth{0pc}
\tablehead{
\colhead{CRGC \#}  &  \colhead{RA}  & \colhead{DEC}  & \colhead{m$_{\rm B}$} &
\colhead{$m_{\rm V}$}  &  \colhead{$m_{\rm R}$} &\colhead{$m_{\rm I}$} & \colhead{B-V}   &
\colhead{V-R\tablenotemark{a}}  & \colhead{V-I\tablenotemark{a}}   &
\colhead{z$_{est}$} & \colhead{z$_{obs}$} \\
\colhead{} & \colhead{(J2000)} & \colhead{(J2000)} & \colhead{(mag)} & 
\colhead{(mag)} & \colhead{(mag)} & \colhead{(mag)} & \colhead{(mag)} & \colhead{(mag)} 
& \colhead{(mag)} & \colhead{}  & \colhead{} }
\startdata
     1     & 02:39:58.2  & -01:36:59  & \nodata    &  22.98    & 
\nodata    & 21.32   & \nodata   & \nodata  & 1.66  & 0.67 & \nodata\\
     2     & 14:11:11.7  & +52:12:01  & \nodata    &  \nodata      & 
20.34   & \nodata     & \nodata   & 1.00  & \nodata   & 0.43 & 0.454\\
     3     & 12:36:57.2  & +62:12:26  & \nodata    &  23.02    & 
\nodata    & \nodata     & \nodata   & \nodata  & \nodata   & 0.68 & 0.561\\
     4     & 02:40:55.1  & -08:22:17  & \nodata    &  \nodata      & 
\nodata    & 21.02   & \nodata   & \nodata  & 1.50   & 0.59 & \nodata\\
     5     & 04:43:11.2  & +02:10:12  & \nodata    &  \nodata      & 
21.62   & \nodata     & \nodata   & 1.00  & \nodata   & 0.61 & 0.532\\
     6     & 21:40:14.0  & -23:40:17  & \nodata    &  \nodata      & 
22.52   & \nodata     & \nodata   & 1.00  & \nodata   & 0.79 & \nodata\\
     7     & 07:50:46.8  & +14:40:46  & \nodata    &  22.72    & 
\nodata    & 21.30   & \nodata   & \nodata  & 1.42  & 0.63 &
0.629\tablenotemark{b}           \\
     8     & 06:11:16.4  & -48:48:29  & 23.67  &  23.45    &   \nodata 
   & 21.74   & 0.22  & \nodata  & 1.71  & 0.76 & \nodata\\
     9     & 03:06:16.3  & +17:20:28  & \nodata    &  \nodata      & 
21.92  & \nodata     & \nodata   & 1.00  & \nodata   & 0.66 & \nodata\\
    10     & 10:07:58.8  & +07:30:09  & \nodata    &  \nodata      & 
22.54  & \nodata     & \nodata   & 1.00  & \nodata   & 0.79 & 0.580\\
    11     & 17:09:59.8  & +10:32:02  & \nodata    &  23.53    & 
\nodata    & \nodata     & \nodata   & \nodata  & \nodata   & 0.78 & 
\nodata\\
    12     & 16:03:09.1  & +42:46:03  & \nodata    &  \nodata      & 
21.37  & \nodata     & \nodata   & 1.00  & \nodata   & 0.57 & \nodata\\
    13     & 04:56:47.2  & +03:52:32  & \nodata    &  \nodata      & 
\nodata    & 21.97   & \nodata   & \nodata  & 1.50   & 0.77 & \nodata\\
    14     & 18:07:01.4  & +45:44:12  & \nodata    &  23.34    & 
\nodata    & 21.73   & \nodata   & \nodata  & 1.61  & 0.74 & \nodata\\
    15     & 10:47:53.3  & -25:14:08  & \nodata    &  \nodata      & 
\nodata    & 22.25   & \nodata   & \nodata  & 1.50   & 0.83 & \nodata\\
    16     & 13:15:22.3  & +49:09:25  & \nodata    &  \nodata      & 
\nodata    & 22.48   & \nodata   & \nodata  & 1.50   & 0.89 & \nodata\\
    17     & 07:27:42.8  & +69:06:47  & \nodata    &  20.64    & 
\nodata    & 19.37   & \nodata   & \nodata  & 1.27  & 0.35 & \nodata\\
    18     & 02:40:57.4  & -08:23:25  & \nodata    &  \nodata      & 
\nodata    & 19.42   & \nodata   & \nodata  & 1.50   & 0.38 & \nodata\\
    19     & 12:30:19.0  & +12:21:54  & \nodata    &  22.94    & 
\nodata    & \nodata     & \nodata   & \nodata  & 1.50   & 0.66 & \nodata\\
    20     & 12:56:57.8  & +47:20:20  & \nodata    &  \nodata      & 
22.04  & \nodata     & \nodata   & 1.00  & \nodata   & 0.68 & 0.996\\
    21     & 02:39:58.4  & -01:36:34  & \nodata    &  22.65    & 
\nodata    & 21.28   & \nodata   & \nodata  & 1.37  & 0.61 & \nodata\\
    22     & 19:38:09.0  & -46:20:48  & \nodata    &  23.59    & 
\nodata    & 21.83   & \nodata   & \nodata  & 1.77  & 0.80 & \nodata\\
    23     & 02:40:55.1  & -08:22:43  & \nodata    &  \nodata      & 
\nodata    & 21.78   & \nodata   & \nodata  & 1.50   & 0.73 & \nodata\\
    24     & 15:06:26.7  & +01:43:11  & \nodata    &  23.81    & 
\nodata    & 21.91   & \nodata   & \nodata  & 1.90  & 0.85 & \nodata\\
    25     & 12:50:02.1  & +39:52:21  & \nodata    &  \nodata      & 
\nodata    & 19.93   & \nodata   & \nodata  & 1.50   & 0.44 & \nodata\\
\enddata

\tablenotetext{a}{Since a $V$-band magnitude is necessary for
the redshift estimation, when no $V$-band image was available,
we have assumed a color of $V$-$R$ = 1.00 or $V$-$I$ = 1.50
to determine the $m_{\rm V}$ from $m_{\rm R}$ or $m_{\rm I}$.
The ``:'' indicates an uncertain redshift.}

\tablenotetext{b}{This redshift is uncertain.}

\end{deluxetable*}

\subsection{Optical Spectroscopy}

Optical spectroscopy for several of our CRGCs was obtained on 21
January 2002 using the Double Spectrograph at the
Palomar\footnote{Observations at the Palomar Observatory were made
as part of a continuing cooperative agreement between Cornell
University and the California Institute of Technology.}
5-m Hale telescope.
Complete spectral coverage from 4000 to 8000 \AA\ was obtained by
using dichroic filter \#55 with a transition wavelength at 5500 \AA,
to split the light into a blue and red beam. In the blue beam, the
600 l/mm grating was used to observe the wavelength range of 4000 to
5800 \AA\, with a dispersion of 1.73 \AA/pixel and a resolution of
$\sim$4 \AA. In the red beam, which covered the wavelength range of
our main interest,
we used the 316 l/mm grating resulting in a dispersion
of 2.44 \AA/pixel over the wavelength range of 5500 to 8000 \AA.
A slit width of 1 arcsec was used which produced spectral
resolution of $\sim$5 \AA.   Both sides of the spectrograph were
equipped with a thinned 1024$\times$1024 Tek CCD.
Each CCD had a gain of 2.0e- per DN 
with a readout noise of 8.6e- for the blue beam and 7.5e- for
the red beam.

The spectra were reduced and analyzed with the IRAF\footnote{IRAF
is distributed by the National Optical Astronomy Observatories,
which are operated by the Association of Universities for Research
in Astronomy,
Inc., under cooperative agreement with the National Science
Foundation.} package. The spectral reduction included bias
subtraction, scattered-light corrections,
and flat-fielding with dome flats.
The two-dimensional images were rectified based on the arc-lamp
observations and the trace of standard stars.
Relative flux calibration was obtained by observations of
standard stars from the list of \cite{Oke90}.
Since the night was non-photometric, only the
standard stars observed contiguous to our targets were used to
generate the sensitivity function.
The results of the spectroscopic observations of our 
four CRGC targets follow.

\subsubsection{CRGC 5}

CRGC 5 lies in close proximity to the x-ray cluster MS0440 + 0210
($z$ = 0.190) and was identified as a possible gravitational arc
(A1) by \citet{Luppino93}.
Spectroscopic observations of A1 by \citet{Gioia98}
produced a redshift of 0.5317, suggesting it may
be weakly lensed by the foreground cluster.  We observed
CRGC 5 for a total exposure time of 3600 seconds.  Several
strong emission lines indicative of significant star formation
in progress,
specifically [O II] $\lambda$3727, H$\beta$,
and [O III] $\lambda$5007, are present (see Figure 2).
These three lines give a redshift of 0.5322 $\pm$ 0.0005,
consistent with the redshift obtained by \citet{Gioia98}.

\subsubsection{CRGC 10}

CRGC 10 was observed for a total of 9000 seconds.  A single
strong emission line at 7679.0 \AA\ is present (see Figure 2).
We tentatively identify this line as H$\beta$ leading to a
redshift of 0.580.  If this H$\beta$ identification is
correct, then other strong emission lines, unfortunately, fall
directly on night sky emission lines.  The line of [O II]
$\lambda$3727 falls atop the strong line of
sodium D at 5890 \AA\ and the line of [O III] $\lambda$5007
falls on the night sky line at 7913.7 \AA.
This night sky line at 7913.7 \AA\ is just a medium strength
emission line.  The resultant sky subtracted spectrum
does show an excess at this wavelength possibly due to the
presence of
[O III] $\lambda$5007 in the spectrum of this galaxy.
The alternative identification of the emission line at
7679.0 \AA\ as
[O II] $\lambda$3727 would result in a galaxy redshift of 1.06.
Our estimated redshift for CRGC 10 (see Section 3.1) is 0.85,
lying between 0.580 and 1.06.
But, given the relative brightness of the apparent ``intruder''
galaxy, it seems more likely that CRGC 10 is an under-luminous
galaxy at $z$ = 0.580 rather than
an over-luminous galaxy at $z$ = 1.06.

\begin{figure}[!ht]
\epsscale{1.1}
\figurenum{2}
\plotone{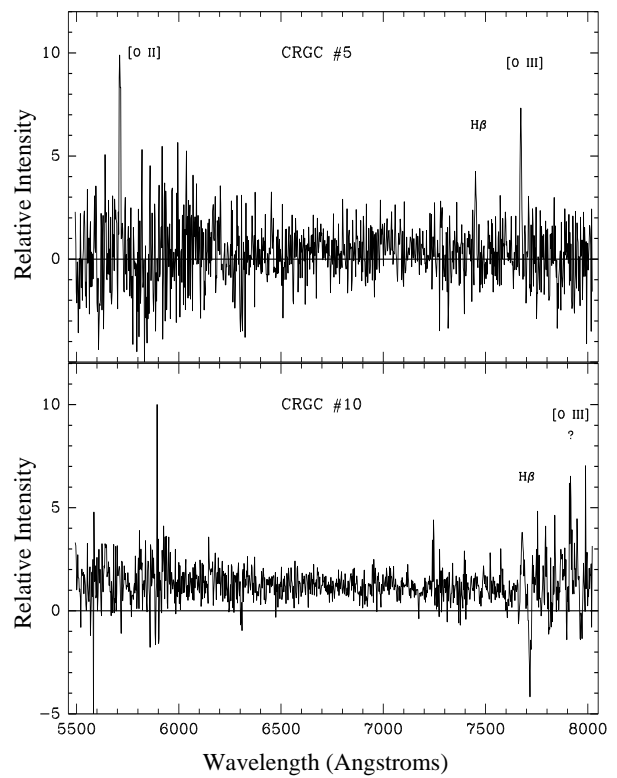}

\caption{ Spectra of CRGC \#5 ($z$ = 0.532) and \#10 ($z$ = 0.580)
obtained with the Palomar 5.0-meter telescope. }

\end{figure}

\subsubsection{CRGC 7 and CRGC 11}

These two CRGCs were observed for 7200 seconds and 4800 seconds, respectively.
For CRGC 7, a weak continuum is present.  An extremely weak emission
line-like feature is present in the 2-dimensional images,
but its presence in the extracted 1-dimensional spectrum is not
easily revealed.
At the pixel location measured in the spectral image, there is
a feature with a peak similar to the larger noise peaks but
with a width unlike the noise peaks in being several pixels in
width.  This feature is of interest as it shows a tilt which may
be indicative of rotation and\/or expansion velocity of the galaxy.
Identifying this weak feature as [O II] $\lambda$3727 leads to
a redshift of 0.629, which is uncertain,
but very close to the estimated redshift of 0.68 for this CRGC.
Lastly, the extracted spectrum of CRGC 11 reveals no information
about this galaxy.

\section{Discussion}

\subsection{Estimated Redshifts for the CRG Candidate Sample}
\label{sec:morph} 

Our goal in identifying this sample of distant CRGCs is to
determine the value of $m$, the exponent used to quantify the
evolution of the galaxy merger/interaction rate with redshift.
In order to accomplish this, either individual redshifts for
our CFGC sample or the redshift interval in which these
CRGCs lay is required.  At this time, measured redshifts are
available for only five of the 25 CRGCs in our sample.
However, we have been able to constrain the redshift interval
of our sample based on estimated redshifts for the remaining
20 CRGCs.

We have used the sample of 11 collisional ring galaxies studied
by \citet{Appleton97} to determine an ``average''
absolute $V$ magnitude, using a value for the Hubble Constant
of $H_{0}$ = 71 km s$^{-1}$ Mpc$^{-1}$.
Extinction corrections and $k$-corrections to the apparent $V$
magnitudes were made by Appleton \& Marston.
The resulting absolute $V$ magnitudes for these galaxies were
averaged to determine a ``standard'' absolute $V$ magnitude of
$M_{V}$ = -21.1 $\pm$ 1.1 for this sample.  While this sample
of CRGs is not a statistically complete sample,
it was selected to be a representative sample.
This value of $M_{V}$ is consistent with an L* galaxy having
several tenths of a magnitude increase in luminosity due to
the star formation associated with the ring structure.
This may also indicate that similar to spiral structure,
the target galaxy must be relatively large for the development of
a coherent ring structure.

We have calculated the expected apparent $V$ magnitude as a
function of redshift for an $M_{V}$ = -21.1 galaxy.
The $k$-correction values used were that of an average Sc galaxy
\citep{Pence76} and we have assumed an average $V$-band galactic
extinction, $A_{V}$, of 0.1.  The cosmological parameters
($H_{0}$, $\Omega_M$, $\Omega_k$, and $\Omega_{\Lambda}$)
and the values used are discussed in more detail in Section 3.2.
The estimated apparent $V$ magnitude as a function of redshift
is shown in Figure 3.  Based on this $m_{V}$-$z$ relation,
estimated redshifts for our sample
of 25 CRG candidates were determined and are given in Table 1.

As part of several unrelated observational programs,
measured redshifts have been published for four of the CRGCs
in our sample:
CRGC 2 at $z$ = 0.454 (3C295 \#122, Dressler \& Gunn 1983),
CRGC 3 at $z$ = 0.561 (HDF 3-773.1, Cohen et al. 1996),
CRGC 5 at $z$ = 0.532 (MS 0440+0204 A1, Gioia et al. 1998), and
CRGC 20 at $z$ = 0.996 (3C 280, Spinrad at al. 1985).
Our spectroscopic observations have confirmed the redshift
for CRGC 5 and added a redshift for a fifth CGRC,
CGRC 10 at $z$ = 0.580.  For this group of five CRGCs,
the agreement between our estimated
redshifts and the measured redshifts is reasonably good
(see Table 1) and
provides support for our assumption that this sample of CRGCs
lie in the redshift interval of 0.1 $\le z \le$ 1.

\begin{figure}
  \epsscale{1.2}
  \figurenum{3}
  \plotone{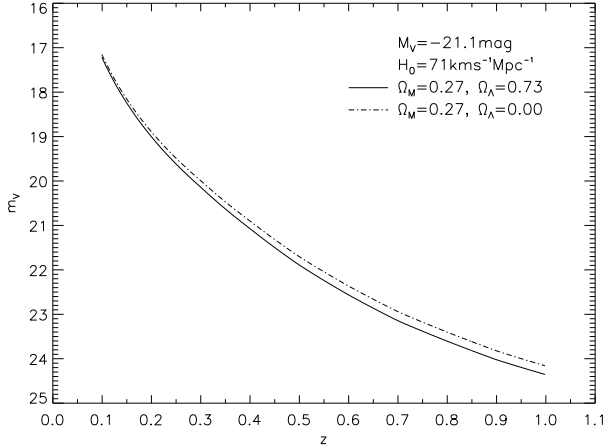}

  \caption{ Apparent $V$ magnitude versus redshift for a ``standard''
    absolute $V$ magnitude ring galaxy for two cosmological models.
The solid curve was used to determine
the estimated redshifts given in Table 1. }

\end{figure}

\subsection{The Evaluation of $m$}

Given that we have been able to constrain the redshift interval in
which our CRGCs lie, it is possible to compare the number of CRGCs we
have identified with that expected for various values of $m$, the
galaxy merger rate exponent.  Using the formalism presented in
\citet{Carroll92}, the co-moving volume element as a
function of redshift is given by

{\setlength\arraycolsep{2pt}}
\begin{eqnarray}
dV ~& = &~ \frac{c}{H_{0} (1+z)^2} ~  \times \nonumber\\       
     &&     \frac{D_{L}^2} 
         {(\Omega_M (1+z)^3 + \Omega_k(1+z)^2 + \Omega_{\Lambda})^{1/2}} 
         ~ d \Omega ~ dz
\end{eqnarray}

where $D_{L}$ is the luminosity distance for a redshift $z$.
In a flat universe, $D_{L}$ can be evaluated using the
integral expression

{\setlength\arraycolsep{2pt}}
\begin{eqnarray}
D_{L} & =& \frac{c(1+z)}{H_{0}}  \times \nonumber\\
      & &   \int_{0}^{z} 
           [ (1+z)^2(1+\Omega_M z) - z(2+z)\Omega_{\Lambda}]^{-1/2} ~ dz.
\end{eqnarray}

(see \citet{Carroll92} for the more general expression).

The parameters in these formulae are $c$, the speed of light,
$H_{0}$, the Hubble constant, $d\Omega$, the solid angle element,
and $dz$, the redshift interval.  The three subscripted
omega terms are used
to parameterize the expansion properties of the Universe where
$\Omega_M$, $\Omega_k$, and $\Omega_{\Lambda}$ are the fractional
contributions due to matter, curvature,
and the cosmological constant (dark energy), respectively.
These three omega terms are related as such,

\begin{equation}
\Omega_M + \Omega_k + \Omega_{\Lambda} = 1
\end{equation}

with $\Omega_0$ = 1 - $\Omega_k$.
Presently, the best determined values for these cosmological
parameters imply a flat ($\Omega_0$ = 1, $\Omega_k$ = 0)
universe with $\Omega_M$ = 0.27, $\Omega_{\Lambda}$ = 0.73,
and $H_{0}$ = 71 km s$^{-1}$ Mpc$^{-1}$ \citep{Bennett03}.

     The total solid angle of sky, $d \Omega$, that we have scanned
is equal to the angular area of 162 WFPC2, each
consisting of three 800$\times$800 CCDs.
The linear image scale for the Wide Field CCDs in the
WFPC2 camera is 0.10 arcseconds per pixel.
We did not include the higher resolution PC field in our scanned
area due to the lower signal-to-noise ratio of the data in this CCD.
Each WF CCD images a solid angle of 1.50 $\times$ 10$^{-7}$ str.
With a total of 486 individual CCD images, the total solid angle,
$d \Omega$, is equal to 7.31 $\times$ 10$^{-5}$ str.
We have not applied any correction to this solid angle,
which would be a reduction of approximately 10\%,
for the area obscured by bright stellar images or the area
on each CCD (approximately 30 rows and 30 columns) affected by the
WFPC2 reimaging mirrors.

     For the local collisional ring galaxy volume density,
we have used the value determined by FM86
of 5.4 $\times$ 10$^{-6}$ $h^3$ Mpc$^{-3}$ (see Section 4.1),
where $h$ = $H_{0}$/100 km s$^{-1}$ Mpc$^{-1}$.
In their determination of the local volume
density, FM86 applied a correction of 20\% to their observed value to
account for edge-on systems which would not be identified as ring
systems due to their inclination.  Therefore, for our estimates of
the expected number of CRGs, we have reduced the FM86 value by 20\%.
This gives a value of 4.3 $\times$ 10$^{-6}$ $h^3$ Mpc$^{-3}$.
We stress here that we have {\em not} introduced any correction
due to our reclassification of the P-type rings (see Section 2.1).

Using equations 1 and 2, we have integrated the product of the
co-moving volume element, the local collisional ring galaxy volume
density, and the galaxy merger rate parameter of $(1 + z)^{m}$ over
the redshift interval of 0.1 $\le z \le 1$.
This produced the expected number of CRGCs in our survey
solid angle for various values of $m$, the results of which are
presented in Table 2.

\begin{deluxetable}{ccc}
\tablecaption{Expected Number of CRGs\label{tbl2}}
\tablewidth{0pc}
\tablehead{
\colhead{Exponent $m$} & \colhead{Number of CRGs} & \colhead{Number of CRGs}\\
 & \colhead{$\Omega_M=0.27$, $\Omega_{\Lambda}=0.73$} & \colhead{$\Omega_M=0.27$, $\Omega_{\Lambda}=0$}}
\startdata
0 &     1.4 &    0.8\\
1 &     2.4 &    1.4\\
2 &     4.1 &    2.5\\
3 &     7.2 &    4.3\\
4 &    12.8 &    7.6\\
5 &    22.9 &   13.6\\
6 &    41.5 &   24.6\\
7 &    75.6 &   44.7\\
8 &   138.8 &   81.9\\
\enddata

\tablecomments{In the above table we present the number of CRGs
expected to be detected in our survey as a function of the merger rate
exponent $m$ for two different sets of cosmological parameters of a
flat universe ($\Omega_k=0$). Note that our results are independent of
the value of the Hubble constant $H_0$ (see Section 3.2).}
\end{deluxetable}

It is interesting to note that since the local collisional
ring galaxy volume density is a function of $H_{0}^3$ and the
co-moving volume element is a function of $H_{0}^{-3}$
{\em our estimate of the number of CRGCs is independent of
the value of $H_{0}$}.
For an $\Omega_M=0.27$, $\Omega_{\Lambda}=0.73$ universe,
our identification
of 25 CRGCs is consistent with an exponent value of $m$ = 5.2,
suggesting a very steep increase in the galaxy merger/interaction
rate with redshift.
Given our relatively large sample, we can assume our Poisson
distribution can be represented by a Gaussian distribution
with 1$\sigma$ error bars being
equal to $N^{0.5}$.  In this case with $N$ = 25,
we find $m$ = 5.2 $\pm$ 0.7.
We rule out a non-evolving ($m$ = 0) and a slowly evolving
($m$ = 1-2) galaxy interaction/merger rate at greater than the
4$\sigma$ level of confidence.

We wish to emphasize two important points.  First, our steeply
increasing galaxy merger rate is not very sensitive to the possible
misclassification of several non-collisional ring galaxies as
collisional rings.  Even if the contamination were as high
as 50\%, the value of $m$ would decrease from 5.2 to 4.0,
still a very steeply increasing galaxy merger rate.
Second, in our determinations for the expected number of
CRGs for various values of $m$, we have used the FM86 value
for the CRG volume density.  Our ``reclassification'' of the
P-type rings in their sample would reduce the local CRG volume
density by a factor of three, decreasing the
expected number of CRGCs in our {\em HST} fields by
the same factor.

     This value for $m$ of 5.2 $\pm$ 0.7 is consistent with
that of \citet{Conselice03}
($m$ $\sim$ 4-6 for luminous galaxies in the {\em HDF}),
but it is significantly higher than that of \citet{Lefevre00}
and the most recent analysis based on galaxy pairs by
\citet{Patton02} of $m$ = 2.3 $\pm$ 0.7.
It is possible that in addition to the galaxy merger rate,
our large value of $m$ may also be indicative of the
evolution of several other galaxy properties.
First, evolution of the types of orbits for the intruder
galaxy may be important.  If the distribution of orbits of
the intruder galaxies has evolved, with low angular
momentum being more common in the past, this would increase
the likelihood of interactions that produce ring galaxies
(C. Struck, private communication).
Second, evolution of the morphological
distribution of the general field population of galaxies,
specifically a higher fraction of gas-rich disk systems
compared to the field today.  Unless
the system being affected by the intruder galaxy is relatively
gas-rich, the interaction will not produce the star formation that
delineates the ring structure.  If many of the nearby elliptical
galaxies are the result of the merger of two disk systems,
the exhaustion of gas in the process of forming these galaxies
will mean that any subsequent interaction will not lead to
enhanced star formation or a ring structure.

\section{Corrections and Potential Biases}

     In order to quantify the evolutionary rate of any
astronomical process, it is necessary to compare local samples
with distant samples.  This is not always straightforward, as
the properties of local samples are not always well-determined,
and non-representative results can occur if any introduced biases
are not taken into account.  In this sections, we evaluate
various corrections and possible biases and their
affect on our result.

\subsection{The Local Volume Density}

    In order to use our sample of distant CRGCs to estimate the
exponent $m$ which is used to parameterize the galaxy
interaction/merger rate, it is necessary to know the local
volume density of collisional ring galaxies.  
In our determination of the value of $m$, we have used the
collisional ring galaxy volume density of FM86.
This local density is based on ring galaxies identified in the
{\em Catalogue of Southern Peculiar Galaxies and Associations}
\citep{Arp87}, an approximately complete sample of 214 objects
out to a survey depth for rings of 10 arc-seconds or larger
of 278 $\times$ $h^{-1}$ Mpc.  This sample is therefore not
magnitude limited, but angular size limited with roughly 10
resolution elements across the smallest galaxies.
The 214 rings from the {\em CPGA} consisted of
both O- and P-type rings.  FM86 classified a random subset of
69 galaxies, selected upon their availability on A-grade
SERC (J) blue-light plates, into O- and P-type (40\% and 60\%,
respectively) rings.
Applying this division to the total sample of 214 galaxies,
they derived a local ring galaxy volume density of
of 5.4 $\times$ 10$^{-6}$ $h^3$ Mpc$^{-3}$.
This value is consistent with the previous, though less rigorous,
determination of \citet{Freeman74} of
7 $\times$ 10$^{-6}$ $h^3$ Mpc$^{-3}$,
but is lower than that determined by \citet{Thompson77},
20 $\times$ 10$^{-6}$ $h^3$ Mpc$^{-3}$.  This larger value is
suspect as the survey fields of \citet{Thompson77}
were located near or on Abell clusters and a number of the
identified objects are quite small and may not actually
be ring galaxies \citep{Appleton97}.

    To be able to compare our distant sample of CRGCs with the
local sample, it is necessary to ensure consistency in the
identification process for both samples.  This was done in
two ways.  First, to ensure that our distant CRGC sample
was identified in approximately the same rest-frame wavelength
range as FM86, who used the SERC J plates for their ring galaxy
classifications,
our search and identification process was done only on those WFPC2
images obtained in the broad red wavelength
filters (F602W, F622W, F704W, and F814W), which would be similar
to the $B$-band in the rest-frame of our sample.
Second, as mentioned earlier, as part of insuring consistency
between classifiers, each classifier looked at a sample of 50 ring
galaxies from the FM86 sample and typed them as either O-type
(resonant) rings or P-type (collisional) rings.  While each
classifier typed the 9 O-rings in the sample as such, many of
the P-rings typed by FM86 were not.  With a very narrow range
of $\pm1$ galaxy, each classifier typed only 14 of
the 41 galaxies as P-rings.

     In our determination of $m$, we have used the local
volume density of FM86, presently the best determined value.
However, based on our re-classifications of the CRGs in the local
sample, we have used much stricter criteria in identifying
the distant sample of CRGCs.  There is a factor of $\sim$3
difference in the number of rings classified as P-type
by FM86 and by us.
If this factor of 3 is applied, thereby lowering the local
collisional ring galaxy volume density by this factor,
the result is a larger value of $m$, with $m$ = 7.0.

\subsection{The Survey Area}

    In our determination of the solid angle covered in our survey,
we have used an image scale of 0.1 arcseconds per pixel with the
dimensions of each of the three low-resolution CCDs being
800 $\times$ 800 pixels.
But, due to the reimaging optics of the WFPC2,
regions along two sides of each CCD are not illuminated by the sky.
The regions are L-shaped strips along the edges of each CCD with
a width of $\sim$30 pixels.  Therefore, the imaging area of each
CCD is actually only 770 $\times$ 770 pixels,
reducing the survey solid angle by $\sim$7.5\%.
In addition, some area was lost due to
bright stars and nearby galaxies, which we estimate to be a few
percent.  Overall, these factors combined lead to a reduction of
the survey solid angle by $\sim$10\%.
Correcting for this lost area leads to an increased value of
$m$ of $\sim$0.15.

\subsection{Incompleteness}

    Incompleteness can enter into and affect our distant CRGC
sample in two ways:  bright CRGCs overlooked in our visual search
and distant CRGCs missed due to being of low signal-to-noise.
Any aspect of incompleteness will result in our underestimating
the number of distant CRGCs, leading to an increased value for $m$.
But, from Table 2, it can be seen that the expected number of
CRGs increases dramatically for large values of $m$, close to a
factor of two for each integer change in the value of $m$.
Therefore, to greatly affect the value of $m$, the level of
incompleteness must be significant.

    Our visual search for CRGCs was done independently by several
different groups of three of the authors in all of the fields.
In the identification of
our distant CRGC sample, there was only 1 CRGC out of the 25
that was identified by a single person.  This was CRGC 12, which
was overlooked due to its relatively high surface brightness.
This occurred early in our search and led to greater care in
examining the full dynamic range of the images.
The remaining 24 CRGCs were identified by at least two searchers,
with the large majority being identified by all three searchers.
For this reason, we feel confident in the thoroughness of
our visual search process.

    The second aspect that contributes to possible incompleteness
are the distant CRGCs not identified due to their
low signal-to-noise in the available images.
To evaluate this incompleteness level, we have divided
our sample of CRGCs into 4 redshift bins to compare it with the
expected redshift distribution based on the combination of
volume and evolutionary rate.  Figure 4 shows the percentage
distribution with estimated redshift determined for our sample
(cross-hatched histogram) along with the expected
distribution (dotted-line histogram) based on the evolutionary
model.
There would appear to be a significant deficit of objects in the
most distant redshift bin.

    While it may be possible to estimate the number of CRGCs
missing from the highest redshift bin, the simplest approach to
account for this incompleteness is to reduce the redshift
interval over which we integrate to determine the value of $m$.
If we disregard the last redshift bin due to incompleteness,
this leads to a reduction of the number of CRGCs from 25 to 21
(the 16\% in the last redshift bin represents 4 CRGCs) out to
a maximum redshift of 0.8 rather than unity.  This leads to a
value of $m$ = 6.8, a significant increase.

\begin{figure}
  \epsscale{1.2}
  \figurenum{4}
  \plotone{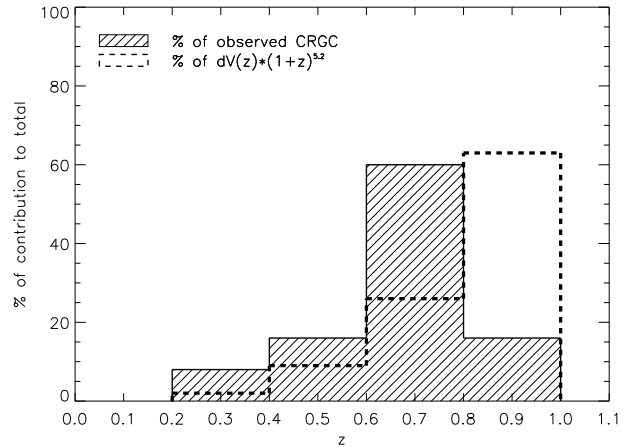}
  
  \caption{Histogram of the percentage distribution of our CRGC
    sample with redshift (hatched histogram) compared with the
    percentage distribution of the product of the co-moving volume
    and evolutionary factor (broken-line histogram).}

\end{figure}

    However, we must qualify this simple approach.  We initially
determined the estimated redshifts for our sample of CRGCs
simply to place a constraint on the redshift range over which
the integration for determining the value of $m$ was conducted,
namely that the redshifts of our sample were less than 1.
But, the errors associated with these estimated redshifts are
relatively large for several reasons.  First, we have assumed
for our distant sample an absolute magnitude of
$M_{v}$ = -21.1, which was determined using a sample of nearby
collisional ring galaxies.  This local sample had a scatter of
$\pm$1 magnitude about this average value, which introduces an
error of 0.1 to 0.15 in the estimated redshifts.
Additionally, the estimated redshifts are quite sensitive to
the assumed k-correction.  While we have used the k-corrections
for an Sc galaxy, the use of the k-corrections for an Scd galaxy
leads to estimated redshifts larger by approximately 0.07 and
increases the fraction of galaxies in the most distant redshift
bin from 16\% to 36\%!  This still indicates some level of
incompleteness in the most distant redshift bin, though
not as severe.  Once again,
excluding the most distant redshift bin leads to a sample size
of 16 CRGCs with $z$ $\le$ 0.8 and a value of $m$ = 6.2.

\subsection{Parallel/Pointed Observations}

    Our sample of 162 WFPC2 fields consists of two very different
types of observations:  parallel fields and pointed observations.
The parallel WFPC2 fields were exposed while other $HST$
instruments were in primary use and should, therefore,
constitute a random
sampling of distant field galaxies.  There are 118 parallel
fields in our total sample.  The remaining 44 fields are
pointed observations towards distant radio galaxies and clusters
of galaxies.  These pointed fields were part of the initial
field sample due to their long exposure times and depth.
Clusters of galaxies are regions of high galaxy density and 
radio galaxies may have an excess of field galaxies associated
with them.  Is it possible the inclusion of these fields has
produced our high value for $m$? 

     Overall, there is no excess of CRGCs in the pointed fields
compared to the parallel fields.  The 118 parallel fields
account for 73\% of our surveyed solid angle.
Based on this percentage,
there should be 18.2 CRGCs in the parallel fields and we have
identified 17 CRGCs, a difference that is not statistically
significant.

     Our WFPC2 sample contains 26 pointed observations towards
radio galaxies.  Observations of the environments of distant
radio galaxies suggest $\sim$25\% of such radio galaxies
are in ``rich'' (Abell richness 0, 1) clusters of galaxies
\citep{Harvanek01,Yates89,Hill91,Zirbel97}.  Inspection of our
radio galaxy fields does not reveal a relatively large
enhancement of field galaxies clustered with the targeted
radio galaxies.  Much more likely is an enhancement in the
field galaxies associated
with the radio galaxies by factors of only 2 to 5,
which would not be obvious by inspection.
But, such an excess would have only a small affect
on our results.
To be associated with the radio galaxy, the field galaxies 
would have to have an almost identical redshift to that of
the radio galaxy, as this relatively small galaxy density
enhancement must have a low velocity dispersion
(200-300 km s$^{-1}$).
We have calculated the volume associated with a redshift interval
of $\pm$0.05 for the redshifts of 0.35 to 0.95 with increments of
0.1 in redshift.
This interval of $\pm$0.05 is equal to a velocity relative to
the radio galaxy of 1500/(1 + $z$) km s$^{-1}$, which is
approximately 3 to 4 times the expected velocity dispersion.
The percentage of the volume associated with $z$ $\pm$ 0.05 to our
total volume ranges from 0.6\% at $z$ = 0.35 to 2\% at $z$ = 0.95.
Given this small fraction of volume with respect to the total
volume along the line-of-sight, enhancements of the field
galaxy density by factors of 2 to 5 would
have minimal consequences.

     Observationally, we find no excess of CRGCs in our radio
galaxy fields.  These 26 fields represent 16\% of our survey
solid angle which leads to an expected number of four CRGCs
in these pointed observations.  We have identified three
CRGCs in these fields, with one of our CRGCs being the
targeted radio galaxy 3C 280 (CRGC 20).  This slight
under-enhancement of CRGCs is not statistically significant.

     Our surveyed sample of WFPC2 fields includes 18 pointed
observations towards distant rich clusters of galaxies, which
are regions of high galaxy density.
However, the environment associated with rich clusters is not
conducive to the production of ring galaxies for two reasons.
First, the galaxy populations in the core regions of these
clusters is predominantly
gas-poor E/S0 galaxies in which the star formation that produces
the ring structure cannot occur.  Second, while the
high galaxy density certainly results in more galaxy
encounters, the velocity dispersions of rich clusters are
also very high ($\sim$1000 km s$^{-1}$).  The relative velocity of
the galaxies in these encounters will be considerably large,
on the order of several times this dispersion, while 
the production of collisional ring galaxies results from
relatively low velocity ($\sim$200-300 km s$^{-1}$) encounters
\citep{Lynds76,Appleton96,Struck99}
The low relative velocity, indicative of bound groups of galaxies
rather than rich clusters, allows the
intruder galaxy to be present for a relatively long period of time,
resulting in a significant gravitational impulse on the
stars in the target galaxy.  For example, the ``grand design''
spiral structure of M51 is thought to have been produced by an
encounter similar to those that produce collisional ring galaxies
except for the fact that the companion to M51 (NGC 5195) has passed
near the edge of the disk rather than through the center of the
disk \citep{Toomre72}.
It is also due to the requirement of low relative
velocities in these encounters to produce ring galaxies that it
is expected these galaxies will merge on a relatively short
time scale, hence the reason for using CRGs in this program.
The high velocity encounters in clusters are of too short
a duration to produce the necessary gravitational affect
on the stars in the target galaxy.

    We have identified five CRGCs in the 18 fields pointed
towards the clusters of galaxies, an area that is 11\% of our
total surveyed area.  Based on this percentage,
the expected number of CRGCs in these fields is only 2.8,
suggestive of only a slight enhancement of CRGCs in these
cluster fields.
Of these five CRGCs, two have measured redshifts, with one
being at the redshift of the cluster
(CRGC 2 with the 3C 295 cluster)
and one being beyond the cluster (CRGC 5) and possibly being
slightly gravitationally lensed by the foreground cluster.

    The large majority of these distant clusters were
observed because they exhibit the ``Butcher-Oemler'' effect,
an excess of blue galaxies compared to local clusters of
similar richness \citep{Butcher78,Butcher85}.  Spectroscopic
observations of the clusters with the largest blue galaxy
fractions ($\sim$20\%) has revealed that only one-half of
the blue galaxies in the fields of these clusters are
actually associated with the
clusters \citep{Lavery86,Dressler92}.
This is consistent with our meager spectroscopic data
(1 of 2 galaxies are associated with the cluster).
Applying the spectroscopic statistics to our
sample of five CGRCs to remove the cluster CRGCs would
decrease our CRGC sample by 2 or 3 galaxies,
leading to a decrease in the value of $m$ by $\sim$0.2.

     It should be noted that despite their possible
association with rich clusters, these CRGCs may still be
indicative of the properties of the field rather than the
cluster environment.
The most likely explanation for the presence of
the blue star-forming galaxies in these rich clusters is
the recent infall of groups, or clouds, of field galaxies
associated with these clusters
\citep{Lavery88,Lavery92b,Ellingson01}.
This scenario is supported by the large velocity dispersion
observed for this blue galaxy
population \citep{Henry87,Dressler92}.

\subsection{Cosmology}    

    At the present time, it seems that the standard Friedman
models of a decelerating universe do not match the
observational evidence, based on Type Ia supernovae,
which suggests rather that the Universe is now in
a stage of acceleration.
Prior to the discovery of the ``dark energy'' contribution
to the expansion of the Universe, the observational
constraints on the ``standard'' cosmological model were
consistent with $\Omega_M=0.27$ and $\Omega_{\Lambda}=0$.
In this cosmological model, while the estimated redshifts
of our CRGC sample are slightly larger (see Figure 3),
the total volume in the redshift interval of
0.1 $\le z \le$ 1 is $\sim$45\% smaller.
Therefore, the expected number of CRGCs increases for
the various values of $m$, as shown in Table 2.
Our observational result would have produced a larger value
of $m$, with $m$ = 6.0.  This value of $m$ is the most
appropriate for comparing our results with previous
determinations for the value
of $m$ which used the $\Omega_{\Lambda}=0$ cosmology.

\subsection{Summary}

Overall, there are several corrections and biases that
affect of determination of $m$.  However, those that would
lead to a lower value of $m$ are corrections of $\sim0.2$,
while those leading to a higher value of $m$ are much larger,
being $\sim$1-2.
This leads us to conclude that our value of $m$ is more
likely a minimum value and could easily be an additive
factor of 1 or 2 larger.
If we make corrections for all the factors above (reduce the
local CRG density by a factor of 3, remove the cluster fields
from the solid angle and remove the CRGCs identified in these
cluster fields, reduce the new solid angle by 10\%, and
use the k-correction of an Scd galaxy which produces the
lowest incompleteness estimate) and correct for
incompleteness as described above,
we find the value of $m$ to be 7.9.

\section{Conclusions}

We have identified a total of 25 collisional ring galaxy
candidates in 162 {\em HST} WFPC2 fields.  This surprisingly
large number of CRGCs implies a galaxy interaction/merger
rate that increases very steeply with redshift.  We find a
minimum value for $m$ of 5.2 $\pm$ 0.7, with $m$ possibly
being as high as 7 or 8.
This large number of distant CRGCs is inconsistent
low values of $m$ (0 $\le m \le$ 2).

Our large value of $m$ may also be influenced by several other
evolutionary effects worthy of future investigation.
First, the frequency of various types of interactions may
be changing, with there being an increase in low angular
momentum, highly radial galaxy collisions as a function
of redshift.  Such collisions are of the type needed for the
production of a collisional ring galaxy.
Second, evolution of the ``target'' galaxy population,
such that gas-rich disk systems, required to sustain an expanding
ring of star formation, would constitute a significantly
higher fraction of the field population at $z$ $\sim$ 1 than
locally.  

\acknowledgments

We thank Michael Reed for his contributions in the initial
undertaking of the project.
We thank the anonymous referee whose comments helped improve
this paper.
VC acknowledges the support of JPL contract 960803.
Some of the data presented in this paper were obtained from
the Multimission Archive at the Space Telescope Science Institute
(MAST). STScI is operated by the Association of Universities for
Research in Astronomy, Inc., under NASA contract NAS5-26555.
Support for MAST for non-HST data is provided by the NASA Office
of Space Science via grant NAG5-7584 and by other
grants and contracts.

\end{document}